\documentclass[a4paper,balancelastpage,nopreprint,floatfix,twoside]{jasatex}
\usepackage[utf8]{inputenc}
\usepackage[T1]{fontenc}
\usepackage{amsfonts,amsmath,amssymb}
\usepackage[squaren]{SIunits}

\newcommand{\w}{\omega}
\newcommand{\etal}{\emph{et al.} }
\newcommand{\derive}[1]{\frac{d{#1}}{dt}}
\newcommand{\deriven}[2]{\frac{d^{#1}#2}{dt^{#1}}}

\newcommand{\imag}[1]{\,\textrm{Im}\left( #1 \right)\,}
\newcommand{\real}[1]{\,\textrm{Re}\left( #1 \right)\,}

\newcommand{\E}[1]{e^{#1 j\omega t}}
\newcommand{\F}[2]{\,\mathbf{F}_{\left(\substack{#1\\#2}\right)}}
\newcommand{\CE}[1]{\,\mathbf{CE}_{#1}}

\newcommand{\gammamin}{\gamma_{\text{min}}}

\begin{document}
\title[Reed-bore interaction in woodwind instruments]
{Interaction of reed and acoustic resonator in clarinet-like systems}
\author{Fabrice \surname{Silva}}
\author{Jean \surname{Kergomard}}
\author{Christophe \surname{Vergez}}
\email[]{{silva, kergomard,vergez}@lma.cnrs-mrs.fr}
\affiliation{Laboratoire de Mécanique et d'Acoustique UPR CNRS 7051,
 13402 Marseille cedex 20, France}
\author{Joël \surname{Gilbert}}
\email[]{joel.gilbert@univ-lemans.fr}
\affiliation{Laboratoire d'Acoustique de l'Université du Maine UMR CNRS 6613, 72085 Le Mans cedex 9, France}
\date{\today}
\begin{abstract}
Sound emergence in clarinetlike instruments is investigated in terms of instability of the static regime. Various models of reed-bore coupling are considered, from the pioneering work of Wilson and Beavers [``Operating modes of the clarinet'', J. Acoust. Soc. Am. \textbf{56}, 653--658 (1974)] to more recent modeling including viscothermal bore losses and vena contracta at the reed inlet. The pressure threshold above which these models may oscillate as well as the frequency of oscillation at threshold are calculated. In addition to Wilson and Beavers' previous conclusions concerning the role of the reed damping in the selection of the register the instrument will play on, the influence of the reed motion induced flow is also emphasized, particularly its effect on playing frequencies, contributing to reduce discrepancies between Wilson and Beavers' experimental results and theory, despite discrepancies still remain concerning the pressure threshold.
Finally, analytical approximations of the oscillating solution based on Fourier series expansion are obtained in the vicinity of the threshold of oscillation. This allows to emphasize the conditions which determine the nature of the bifurcation (direct or inverse) through which the note may emerge, with therefore important consequences on the musical playing performances.
\end{abstract}


\maketitle

\section{Introduction}
\par Sound production in reed wind musical instruments is the result of self-sustained reed oscillation. The mechanical oscillator, the reed, acts as a valve which modulates the air flow entering into the instrument, by opening and closing a narrow slit defined between the tip of the reed itself and the lay of the mouthpiece. The phenomenon thus belongs to the class of flow-induced vibrations, which has been extensively studied both theoretically and experimentally (see, for example, Blevins\cite{blevins:2001}). Various regimes can occur in such systems: static regime, periodic oscillating regimes, and even complex chaotic behaviors.
\par A first step to study this kind of oscillator, is to analyze the stability of the trivial solution, the equilibrium position of the reed, in order to find a threshold of instability associated with a set of control parameters defining the embouchure (reed, mouthpiece, and player) and the instrument itself. As an output of the threshold analysis, a mouth pressure threshold can be found. It is relevant for the musical playing performance: An estimation of the threshold value is a first evaluation of the ease of playing. In this paper the threshold of instability of the reed of a clarinetlike system is firstly investigated theoretically by revisiting the important pioneering work of Wilson and Beavers\cite{wilson:1974}. Second, small oscillations beyond the instability threshold are considered.
\par To answer the question whether the reed equilibrium is stable or not, several theoretical methods are available and have been used to study reed wind instruments. Using the feedback loop analogy, it is known as the free oscillation linear stability problem in a closed loop obtained when the nonlinear component of the loop is linearized around the trivial solution. Then the linear stability of the reed can be studied with respect to each resonance of the input impedance of the resonator in frequency domain (see, for example, Refs.~~\onlinecite{backus:1963,benade:1968a,fletcher:1991,saneyoshi:1987,chang:1994} and~~ \onlinecite{tarnopolsky:2000} for reed instruments, Ref.~~\onlinecite{elliott:1982} for lip-reed instruments and Ref.~~\onlinecite{fletcher:1993} for a generic type of reeds). Using the dynamical system representation where each resonance of the resonator is described as a simple second order oscillator in time domain, the linear stability analysis requires to solve eigenvalue problems and to analyze the sign of the eigenvalues real part (see, for example, Ref.~~\onlinecite{cullen:2000} for lip-reed instruments and Ref.~~\onlinecite{ruty:these} for vocal folds). The theoretical results can then be compared with experimental ones coming from artificial mouths by playing them as gently as possible (see, for example, Refs.~~\onlinecite{backus:1963,wilson:1974,dalmont:1995} and~~\onlinecite{dalmont:2007} for reed instruments; Ref.~~\onlinecite{cullen:2000} for lip reed instruments; and Refs.~~\onlinecite{steinecke:1995,lopez:2006,ruty:these} and~~\onlinecite{chan:2006} for vibrating vocal folds). The first attempt to derive theoretically the spectrum of reed instruments beyond the instability threshold is due to Worman\cite{worman:these}. His results were at the origin of several works such as Refs.~~\onlinecite{benade:1988,grand:1997,kergomard:2000b}.
\par Despite a rather simple description of physical phenomenon, pioneer theoretical and experimental results concerning mouth pressure and frequency at oscillation threshold were obtained by Wilson and Beavers\cite{wilson:1974}, showing the important role of reed damping determining the clarinetlike or lingual pipe organlike behaviour.
\par The theory of Wilson and Beavers (hereafter denoted WB) is presented and discussed in Sec.~\ref{sec:WBtheory}. Then, model improvements from literature are added in the theory in order to try to reduce discrepancies between WB experiments and the theoretical results (Sec.~\ref{sec:improvements}). Section~\ref{sec:beyondthreshold} is devoted to a study of small oscillations beyond the instability threshold according to the direct bifurcation behavior of the clarinetlike instruments. Finally perspectives are discussed in Sec.~\ref{sec:conlusion}.

\section{Wilson and Beavers theory\label{sec:WBtheory}}
The model and the principal results of WB are reminded. Particular attention is brought to the numerical estimation of thresholds when the frequency approaches the reed resonance frequency, where some discrepancies with WB results are found. Finally the effect of the reed damping on playability is emphasized.

\subsection{Basic physical model}
The physical model used by WB is reviewed with some comments related to more recent literature. It is based on the description of three separate elements: the reed, the bore, and the airflow. The model used here is classical and extremely simplified, but is also proven to be efficient in order to reproduce self sustained oscillations (for sound synthesis examples, see Refs.~~\onlinecite{schumacher:1981} and~~\onlinecite{guillemain:2005}). Additional elements will be discussed in the next section.

\subsubsection{Reed}
\par Based on the fact that reed displacement occurs in the vertical direction mainly without torsion, WB, among many authors, assumed a single degree of freedom motion. Reed-lip-mouthpiece system is thus modeled as a lumped second-order mechanical oscillator with stiffness per unit area $K$, damping parameter $q_r$, and natural angular frequency $\w_r$, driven by the pressure drop $P_m-p(t)$ across the reed, with an inward striking behavior:
\begin{equation}
\deriven{2}{y}+q_r\w_r \derive{y}+\w_r^2\left(y(t)-y_0\right)
=\frac{\w_r^2}{K}\left(p(t)-P_m\right),
\end{equation}
$p(t)$, $P_m$, $y(t)$, and $y_0$, being the mouthpiece pressure, the blowing pressure, the tip opening [denoted $a(t)$ in WB's paper\cite{wilson:1974}], and the tip opening without any pressure difference, respectively. $P_m$ is assumed to be constant.
\par Avanzini \etal\cite{avanzini:2004} numerically showed that this lumped model is valid for a small vibration theory where only the interaction between bore resonances and the first flexural mode of the reed is investigated. Measured transfer functions of a reed mounted on a mouthpiece also show a two degree of freedom response\cite{gazengel:2007}.

\subsubsection{Bore}
\par The behavior of the acoustical resonator is determined by an input impedance relationship between acoustic quantities in the mouthpiece [acoustic pressure $p(t)$ and volume flow $u(t)$, or $P(\w)$ and $U(\w)$, respectively, in the frequency domain]. WB assumed, for a cylindrical bore representing a simplified clarinet body, an expression given by Backus\cite{backus:1963}:
\begin{equation}
\mathcal{Z}_e(\w)=\frac{P(\w)}{U(\w)} = jZ_c\frac{1}{\displaystyle 1-\frac{j}{2Q}}
\tan{\left(\frac{\w L}{c}\left(1-\frac{j}{2Q}\right)\right)},
\label{eq:ZeWB}
\end{equation}
where $j^2=-1$. $c$, $\rho$, $L$, $S$ and $Z_c=\rho c/S$ are the wave speed in free space, density of air, bore length, bore cross section, and characteristic impedance, respectively. The quality factor $Q$ is assumed by WB to be frequency independent, implying a damping  proportional to the frequency.
\par As it will be seen later, this assumption can be discussed, and improved models will be used, because pressure thresholds are strongly influenced by bore losses (which are directly linked to the value of parameter $Q$).

\subsubsection{Airflow}
As noted by Hirschberg~\cite{hirschberg:1995}, in the case of clarinetlike instruments, the control of the volume flow by the reed position is due to the existence of a turbulent jet. Indeed, a jet is supposed to form in the mouthpiece (pressure $p_\mathrm{jet}$) after flow separation from the walls at the end of the (very short) reed channel. Neglecting the velocity of air flow in the mouth compared to the jet velocity $v_\mathrm{jet}$ and assuming a downward air flow ($v_\mathrm{jet}>0$), the Bernoulli theorem applied between the mouth and the reed channel~\cite{hirschberg:1995} leads to
\begin{equation}
P_m = p_\mathrm{jet}+\frac{1}{2}\rho v_\mathrm{jet}^2.
\end{equation}
Assuming a rectangular aperture of width $W$ and height $y(t)$, the volume flow $u$ across the reed channel can be expressed as follows:
\begin{equation}
u(t)=Wy(t)\sqrt{\frac{2}{\rho}}\sqrt{P_m-p_\mathrm{jet}(t)}.
\label{eq:debitcomplet}
\end{equation}
Since the cross section of the mouthpiece is large compared to the cross section of the reed channel, it can be assumed that all the kinetic energy of the jet is dissipated through turbulence with no pressure recovery (like in the case of a free jet). Therefore, pressure in the jet is (assuming pressure continuity) the acoustic pressure $p(t)$ imposed by the resonator response to the incoming volume flow $u$. This model is corroborated by experiments\cite{dalmont:2003}. Similar descriptions are used for double-reed instruments\cite{almeida:2007} and buzzing lips\cite{chick:2006}.

\subsection{Characteristic equation and instability threshold}
The conditions for which self-sustained oscillations become possible are sought, that is, the minimum value of blowing pressure required for the static regime to be unstable is investigated for a given configuration of the experiment ($\w_r$, $q_r$, and $L$ being fixed). Common methods of linear stability analysis\cite{hale:1991} are used in this study, and solutions having time dependence $\exp{(j\w t)}$ are sought. Cancellation of the imaginary part of $\w$ corresponds to an oscillation that is neither damped nor amplified: it characterizes the instability threshold of static regime. Attention is drawn to the fact that this quantity may differ from the oscillation threshold depending on the nature of the bifurcation, studied in Sec.~\ref{sec:harmbal}. As a language abuse \emph{oscillation threshold} is often used instead of \emph{instability threshold}.
\par Assuming small vibrations around equilibrium state (mean values of $y$ and $p$ are $y_0-P_m/K$ and $0$, respectively), the volume flow relationship (\ref{eq:debitcomplet}) is linearized. Dimensionless quantities are introduced here: $\theta$, $\mathcal{Y}_e$, and $D$ are the dimensionless frequency, input admittance and the reed transfer function, respectively.
\begin{equation}
\theta = \displaystyle \frac{\w}{\w_r},\;
\mathcal{Y}_e(\theta ) = \displaystyle\frac{Z_c}{\mathcal{Z}_e(\theta)}\text{ and }
D(\theta ) = \frac{1}{1+jq_r\theta-\theta^2}.
\label{eq:dimensionless}
\end{equation}
There are two dimensionless control parameters: 
\begin{equation}
\gamma = \displaystyle \frac{P_m}{Ky_0}\text{ and }
\zeta = Z_cW \sqrt{\displaystyle \frac{2y_0}{K\rho}}.
\end{equation}
$\gamma$ is the ratio between mouth pressure and the pressure required to completely close the reed channel in static regime,  while $\zeta$ mainly depends on mouthpiece construction and lip stress on the reed and is linked to the maximum flow through the reed channel ($\zeta$ equals quantity $2\beta$ in Ref.~~\onlinecite{wilson:1974}).

Linearization of Eq.~(\ref{eq:debitcomplet}) leads to the so-called characteristic equation:
\begin{equation}
\mathcal{Y}_e(\theta)=\zeta\sqrt{\gamma}\left\{D(\theta )-\frac{1-\gamma}{2\gamma}\right\},
\label{eq:careq}
\end{equation}
which can be split into real and imaginary parts:
\begin{gather}
\imag{\mathcal{Y}_e(\theta)}=\zeta\sqrt{\gamma}\;\imag{D(\theta)},
\label{eq:sys1a} \\
\real{\mathcal{Y}_e(\theta)}=\zeta\sqrt{\gamma}\left(\real{D(\theta)}-\frac{1-\gamma}{2\gamma}\right).
\label{eq:sys1b}
\end{gather}
At last, a dimensionless length $k_r L=\w_r L/c$ is introduced.

\subsection{Numerical techniques}
The unknowns $\theta,\,\gamma\in\mathbb{R^+}$ satisfying Eq.~(\ref{eq:careq}) are numerically determined for a range of bore lengths, parameters $(q_r, \zeta, \w_r)$ being set. They correspond to frequency and mouth pressure at instability threshold of the static regime. When various solutions exist for a given configuration due to the interaction of the reed resonance with the several bore resonances, the threshold observed experimentally by increasing the blowing pressure is the one having the minimum value of $\gamma$.

\par The characteristic equation is transcendental and may have an infinite number of solutions. Zero finding is done using the Powell hybrid method\cite{powell:1970}, which combines the advantages of both Newton method and scaled gradient one. A continuation technique is adopted to provide an initial value to the algorithm: the first resolution is done for very high values of $L$ ($k_r L\simeq\!\! 30$), i.e., for a nearly non resonant reed (i.e., with neither mass nor damping, hereafter denoted the massless reed case), where $f\simeq (2n-1)c/4L$ and $\gamma\simeq 1/3$ (with $n\in \mathbb{N}^*$). Bore length is then progressively decreased and zero finding for a given value of $L$ is initialized with the pair $(\theta, \gamma)$ solution of the previous solving (bore slightly longer). Depending on the initialization of the first resolution ($k_r L =30$), it is possible to explore the branches associated with the successive resonances of the bore.
When reed resonance and bore antiresonance get closer to each other ($k_r L\rightarrow n\pi (n\in \mathbb{N})$), fast variation of the pressure threshold requires adjustment of the step size.

\subsection{Results}
\par
\begin{figure}
\includegraphics[width=\columnwidth]{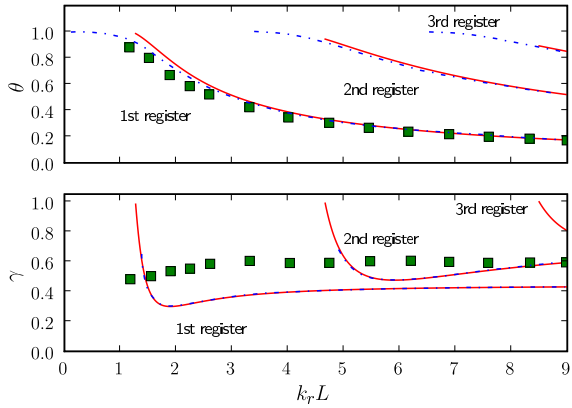}
\caption{Dimensionless threshold frequencies (top) and pressure (bottom) for a strongly damped reed: $q_r=0.4$, $f_r=750\,\hertz$, $\beta=0.065$. Results (dashed lines) and measurements (squares) from Ref.~~\onlinecite{wilson:1974}; our numerical results (solid lines).}
\label{fig:amortfort1}
\end{figure}

\begin{figure}
\includegraphics[width=\columnwidth]{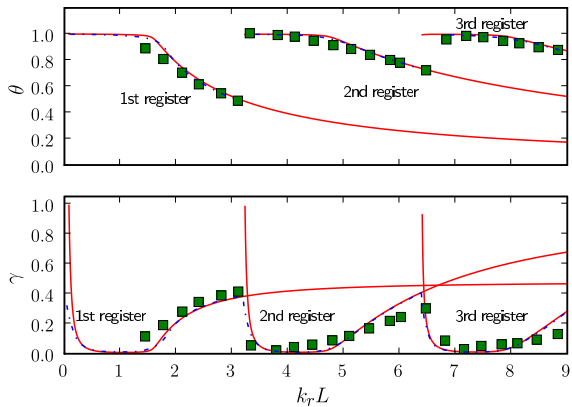}
\caption{Dimensionless threshold frequencies and pressure for a lightly damped reed: $q_r=0.008$, $f_r=700\,\,\hertz$, $\beta=0.05$. Results (dashed lines) and measurements (squares) from Ref.~~\onlinecite{wilson:1974}; our numerical results (solid lines).}
\label{fig:amortfaible1}
\end{figure}

 Two kinds of behavior can be distinguished. For strongly damped reeds (Fig.~\ref{fig:amortfort1}), the threshold frequencies (dashed lines, result from Ref.~~\onlinecite{wilson:1974}) always lie either near the reed resonance ($\theta=1$) or near the first impedance peak frequency of the pipe (hyperbola $\theta=\pi/(2k_r L)$, not represented in the figure for readability), corresponding to the first register of the instrument. When the length $L$ decreases, pressure threshold gradually reduces from values assumed for the massless reed model to a minimum point for $k_r L\sim \pi/2$, and then strongly increases as the pipe becomes shorter. 
When increasing $\gamma$ from 0, the loss of stability of the static regime may give rise to an oscillating solution which always corresponds to the first register since the instability thresholds of the higher registers occur for higher values of mouth pressure.
On the contrary, considering now slightly damped reeds (Fig.~\ref{fig:amortfaible1}), emerging oscillations can occur near higher pipe resonances. Indeed, for certain ranges of $L$, the pressure threshold associated with one particular higher-order register is lower than the pressure required to drive the air column in the first register. This lowest pressure threshold is associated with the acoustic mode, the natural frequency of which being the closest to the reed resonance.


\par These results show the influence of reed damping on selecting
\begin{itemize}
\item a clarinetlike behavior with heavily damped cane reed: preference is given to the \emph{chalumeau} register, i.e., the lowest register, or
\item a lingual organ pipe behavior with very lightly damped metallic reed: tuning is performed by adjusting reed vibrating length by means of a wire adjusted on reed, see Ref.~~\onlinecite{miklos:2006} for further details.
\end{itemize}
This noticeable conclusion of Wilson and Beavers'' paper gainsays von~Helmholtz\cite{helmholtz:1954}, who stated that the differences of behavior are linked to the mass of the reed.

\subsection{Discussion}
Despite the important results obtained by WB, discrepancies concerning numerical results and model limits can be pointed out. In a validation phase of our numerical algorithms, our results (see Fig.~\ref{fig:amortfort1} and~\ref{fig:amortfaible1}) were compared to their ones (the data being extracted from their article), solving exactly the same equation. Differences appear between numerical results concerning threshold frequencies, approaching reed resonance when the length decreases. Investigations pointed out model limits the authors do not take into account: linearization of flow relationship is valid only while reed channel is not closed at rest, i.e., when $P_m<Ky_0$. Using the linear form for higher values of $P_m$ would be meaningless, even for free reed aerophones: the opening function (linked to the reed displacement) taking part in flow calculation can never be negative. For instruments for which reed beats against the mouthpiece, reed channel is completely closed and then sustained oscillations cannot occur for bore lengths where WB theory predicts pressure threshold above static beating reed pressure by extending linearization beyond model limits.
\par The fact that the dimensionless mouth pressure $\gamma$ remains lower than 1 implies the existence of a maximum threshold frequency for the oscillation. We assume that the maximum value $\theta_{\text{max}}$ is associated with the maximum value of $\gamma$ (i.e.,1). No theoretical proof is given here, but this assumption is observed in all numerical results obtained. Then the solution $\theta_{\text{max}}$ is given by
\begin{equation}
\mathcal{Y}_e(\theta_{\text{max}}) = \zeta D(\theta_{\text{max}}),
\end{equation}
which leads to relation
\begin{equation}
1-\theta_{\text{max}}^2 = \zeta \real{\mathcal{Z}_e(\theta_{\text{max}})} >0
\end{equation}
as the bore is a passive system. As a conclusion, it is not possible to play sharper than a frequency slightly flatter than the reed-lip-mouthpiece system resonance frequency, the deviation to this latter being linked to the losses in the bore.

\subsection{Minimum pressure threshold: Improved playability for interacting resonances}
\par For each value of $q_r$, there exists one or more ranges of bore lengths where playability is greatly improved. Indeed, pressure threshold curves show a minimum for a certain value of $k_r L$, denoting an increased easiness to produce the note corresponding to this length. Associating a clarinet mouthpiece with a trombone slide, informal experiments confirm that it is easier to produce some notes than other ones. Analytical approximated expressions for the minimum have been investigated. Under the assumption that this minimal value is obtained for an emerging frequency located close to the reed resonance, and therefore is mainly determined by reed damping, bore losses can be ignored $\real{\mathcal{Y}_e(\w)}=0$, Eq.~(\ref{eq:sys1b}) leading thus to:
\begin{equation}
\gamma = \frac{1}{1+2\real{D(\theta)}}.
\label{eq:gammaRe}
\end{equation}
In the massless reed model ($D(\theta)=1$), threshold pressure is equal to $1/3$ and frequencies correspond to frequencies for which the imaginary part of the bore input impedance vanishes, which is consistent with results already published\cite{kergomard:1995}. The minimum pressure threshold occurs at a maximum of $\real{D(\theta)}$:
\begin{equation}
\real{D(\theta)} = \frac{1-\theta^2}{(1-\theta^2)^2+(q_r\theta)^2},
\end{equation}
obtained for $\theta=\sqrt{1-q_r}$ (which is consistent with the approximation $\theta\simeq 1$), thus,
\begin{equation}
\gamma_0 = \frac{q_r(2-q_r)}{2+q_r(2-q_r)},
\end{equation}
nearly proportional to $q_r$ for lightly damped reeds.
\par In order to evaluate the effect of bore losses on this minimum pressure threshold, a one-mode resonator with losses is now considered:
\begin{equation}
\mathcal{Y}_e(\w)=Y_n\left(
    1+jQ_n\left(\frac{\w}{\w_n}-\frac{\w_n}{\w}\right)\right),
\label{eq:Ye1mode}
\end{equation}
where $Y_n$ is the minimum amplitude of the admittance and $Q_n$ is the quality factor; Eqs.~(\ref{eq:sys1a}) and~(\ref{eq:sys1b}) become
\begin{gather}
\label{eq:annrealce}
Y_n+\zeta\frac{1-\gamma}{2\sqrt{\gamma}}
=\zeta\sqrt{\gamma}\frac{1-\theta^2}{(1-\theta^2)^2+(q_r\theta)^2},\\
\label{eq:annimagce}
Y_nQ_n\left(\frac{\theta}{\theta_n}-\frac{\theta_n}{\theta}\right)
=-\zeta\sqrt{\gamma}\frac{q_r\theta}{(1-\theta^2)^2+(q_r\theta)^2}.
\end{gather}
Ignoring the variation of $Y_n$ with the length of the bore, the derivation of Eq.~(\ref{eq:annrealce}) with respect to $k_r L$ leads to a minimum value of the function $\gamma=f(k_r L)$ for $\theta_{\text{min}}^2=1-q_r$ and $\gammamin$ solution of equation
\begin{equation}
\frac{Y_n}{\zeta\sqrt{\gammamin}}+\frac{1-\gammamin}{2\gammamin}=\frac{1}{q_r(2-q_r)},
\end{equation}
which the first-order solution is given by
\begin{equation}
\gamma_{\mathrm{min}} \simeq \gamma_0
    \left(1+2\frac{Y_n}{\zeta}\sqrt{\gamma_0}\right)
\end{equation}
obtained for
\begin{equation}
\w_n=\w_r \left(1-\frac{q_r}{2}+\frac{1}{2Q_n}+\frac{\zeta}{2Y_nQ_n\sqrt{q_r}}\right).
\end{equation}
For an open/closed cylinder $\w_n=(2n-1)\pi c/2L$, the result is
\begin{equation}
(k_r L)_{\mathrm{min}}\simeq (2n-1)\frac{\pi}{2}
\left(1+\frac{q_r}{2}-\frac{1}{2Q_n}-\frac{\zeta}{2Y_n Q_n\sqrt{q_r}}\right).
\end{equation}
Typical values $Y_n=1/25$, $\zeta=0.4$, and $q_r=0.4$ lead to an increase in $\gamma_{\mathrm{min}}$ relative to  $\gamma_0$ of about $8\%$, confirming the preponderant effect of reed damping on the minimum pressure threshold.
\par In order to understand how coupling acoustical and mechanical resonances could reduce pressure threshold over a wider bore length range, the neighborhood of the previously mentioned minimum has been studied. Using again a single acoustical mode for the calculation, derivation of a parabolic approximation was possible. Writing $\gamma=\gammamin(1+\varepsilon^2)$, $\theta^2=\theta_{\text{min}}^2+\delta$ and $\w_n=(\w_n)_{\text{min}}(1+\nu)$, with $\varepsilon$, $\delta$, and $\nu$ small quantities, the Taylor expansion of Eqs.~(\ref{eq:annrealce}) and~(\ref{eq:annimagce}) with respect to these values leads to the next relationships:
\begin{gather}
\varepsilon^2\sim \delta^2/(2q_r^2),\\
2q_r^2Y_nQ_n(\delta-2\nu)=-\delta \zeta\sqrt{q_r}.
\end{gather}
Finally, near the minimum pressure threshold, dependence on bore length is given by
\begin{equation}
\gamma=\gammamin\left(1+\frac{2 q_r}{\displaystyle \left(q_r^{3/2}+\frac{\zeta}{2Y_n Q_n}\right)^2}
\left(\frac{k_r L-k_r L_{\text{min}}}{k_r L_{\text{min}}}\right)^2\right).
\end{equation}
In a first approximation, the aperture of the approximated parabola, thus the width of the range for which oscillation threshold is lowered, is mainly controlled by the musician embouchure, i.e., by reed damping and lip stress on the reed. This means that, thanks to its embouchure, the player can expect an easier production of tones for certain notes.
\par For a lossy cylindrical open/closed bore, modal expansion of input impedance gives $Y_n Q_n=\w_n L/2c=(2n-1)\pi/4$ so that bore losses do not seem to have a great influence on playing facility, at least when considering minimal blowing pressure $\gamma_{\mathrm{min}}$ ($Q_n$ does not appear alone in first order calculation). On the contrary, they are essential for understanding the extinction threshold phenomenon\cite{dalmont:2005}, i.e., when the reed is held motionlessly against the lay.

\section{Model improvements\label{sec:improvements}}
Last four decades have been fruitful in physical modeling of musical instruments, especially for single reed instruments. Pipes have been the focus of a great number of studies since Benade\cite{benade:1976}, as well as the description of peculiarities of the flow (Backus\cite{backus:1963}, Hirschberg\cite{hirschberg:1995}, and Dalmont \etal\cite{dalmont:2003}). The aim here is to try to reduce discrepancies between WB experiments and theory, based on some of those investigations which look relevant to the study of oscillation threshold. 
In a real clarinet-player system, the vocal tract may have an effect on the oscillation threshold (see, e.g., Ref.~~\onlinecite{fritz:these}). However, the use of porous material in the WB apparatus suggests that acoustic interaction between the chamber and the mouthpiece has been canceled or at least strongly reduced. Therefore we do not discuss the effect of the upstream part.

\subsection{Viscothermal losses model and vena contracta\label{sec:highL}}
\par It should be noticed that there is a gap in pressure threshold values between experiment and theory in WB article. This occurs even for long bores when reed dynamical behavior should not deviate from the ideal spring model (because of an emerging frequency much smaller than $\w_r$). For that case, Kergomard \etal\cite{kergomard:2000b} provided an approximated formula taking into account both reed dynamics and bore losses that can be extended to
\begin{equation}
\gamma \simeq \frac{1-\theta_n^2}{3-\theta_n^2}
+\frac{2\real{\mathcal{Y}_e(\theta_n)}}{3\sqrt{3}\zeta},
\quad \theta_n = (2n-1)\frac{\pi}{2k_r L}.
\label{eq:highL}
\end{equation}
$\theta_n$ corresponding to the $n$th resonance frequency of the bore. In comparison with ideal model (lossless bore and massless reed, i.e., $\gamma=1/3$), additional corrective terms are considered in Eq.~(\ref{eq:highL}), one lowering pressure threshold due to the collaboration of the resonant reed, the other one requiring higher blowing pressure due to dissipation in the bore. According to this approximated expression, pressure thresholds depend on the mouthpiece parameter and on bore dissipation at playing frequency.
\par Now focus is done on using realistic values of $\zeta$ and $\mathcal{Y}_e(\theta)$, assuming acoustic losses in clarinetlike bore to be due mainly to viscothermal dissipation (Ref.~~\onlinecite{benade:1976}). Other kinds of losses such as nonlinearity localized at the open end of a tube are negligible since study is done at oscillation threshold, i.e., for very small amplitude oscillations. Simpler model (``Raman's model'') has been recently investigated by Dalmont \etal\cite{dalmont:2005} and led to satisfactory oscillation threshold of the fundamental register of the clarinet when operating frequencies are much lower than the reed natural frequency (the reed being considered as an ideal spring). In the present study, the magnitudes of the higher order impedance peaks need to be correctly estimated so that neither the loss model of Raman nor that of Backus\cite{backus:1963} may be realistic enough. Pressure thresholds would be inaccurate when higher-order modes of bore oscillate first.
\par The standard formula for the input impedance will be hence considered:
\begin{eqnarray}
\mathcal{Z}_e(\w) &=& jZ_c\tan{(kL)}\text{ with }jk=\frac{j\w}{c}+\eta\sqrt{\frac{j\w}{c}}
\label{eq:kirchhoff}
\end{eqnarray}
where $\eta$ is a coefficient quantifying the viscothermal boundary layers, equals to $0.0421$ in $\meter\kilo\second$ units for a $7\,\milli\meter$ radius cylinder. This model introduces dissipation and dispersion and leads to a zero valued impedance at zero frequency, which is still consistent with the linearization of flow relationship with a zero mean value of acoustic pressure. In this expression, viscothermal effects are ignored in the characteristic impedance (see Ref.~~\onlinecite{kergomard:1995}). Equation~(\ref{eq:kirchhoff}) leads to peak magnitudes decreasing with length $L$, whereas they are not sensitive to $L$ in Backus empirical expression. A direct consequence is that the pressure threshold increases as the resonator lengthens.
\par Another effect may occur and modify pressure threshold. Hirschberg\cite{hirschberg:1995} and Hirschberg \etal\cite{Hirschberg:1990} brought to attention on vena contracta phenomenon: due to sharpness of edges, flow separation may result in the formation of a free jet in reed channel, this contraction effect resulting in a jet cross-sectional area smaller than the reed channel opening. Recent investigations~\cite{dasilva:2007} applying the lattice Boltzmann method to the reed channel confirm previous experiments\cite{vanzon:1990}. The assumption of constant vena contracta is valid in some specific cases, for short channel geometry, but there is no proof yet that this effect is significant for a real clarinet mouthpiece. Here small vibrations of the reed near oscillation threshold are considered, which may induce little influence of the flow unsteadiness on the measured pressure threshold. So it is possible\cite{hirschberg:1995} to include vena contracta phenomenon by reducing the area of the reed channel $Wy(t)$ by a coefficient approximately $0.6$, i.e., by multiplying $\zeta$ by this coefficient.
\begin{figure}
\includegraphics[width=\columnwidth]{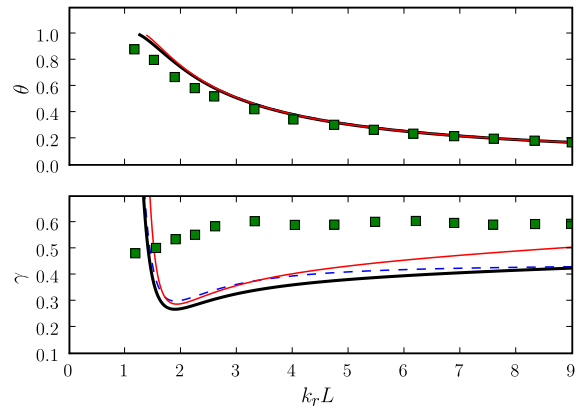}%
\caption{Results with the model considering vena contracta phenomenon (thin line) and viscothermal losses (thick line) (same conditions as in Fig.~\ref{fig:amortfort1}). WB results are also recalled (dashed line).}
\label{fig:venaKirch}
\end{figure}
\par Numerical investigations point out (see Fig.~\ref{fig:venaKirch}) that taking into account realistic losses and vena contracta phenomenon reduces discrepancies in pressure threshold, especially for high values of bore length and for a strongly damped reed, i.e., when reed dynamics has a small influence on oscillation threshold. Nevertheless pressure values experimentally obtained by WB still remain quite higher than the ones corresponding to the modified model, while frequencies are unaltered.

\par An attempt to explain the discrepancies in threshold measurements will be provided in Sec.~\ref{sec:beyondthreshold}.

\subsection{Reed motion induced flow}
The influence of the reed is not limited to its resonance. Since Nederveen\cite{nederveen:1998} and Thompson\cite{thompson:1979}, it is proved that the flow entering through the reed opening is divided in one part exciting the resonator and another part induced by the reed motion. In fact, the vibration of the surface of the reed produces an additional oscillating flow. Thus the entering flow $U$ can be written as
\begin{equation}
U(\w) = \mathcal{Y}_e(\w)P(\w) + S_r (j\w Y(\w)),
\label{eq:debitanche}
\end{equation}
where $S_r$ is the effective area of the vibrating reed related to the tip displacement $y(t)$. Alternately, a length $\Delta l$ can be associated with the fictitious volume where the reed swings. Dalmont \etal\cite{dalmont:1995} reported typical values of $10\,\milli\meter$ for a clarinet. Nederveen\cite{nederveen:1998} linked $\Delta l$ to reed strength (or \emph{hardness}): $\Delta l$ may approximately vary from $6\,\milli\meter$ (strong reeds) to $9\,\milli\meter$ (softer reeds). These values being small compared to clarinet dimensions, the reed motion induced flow can be considered through a mere length correction in common work, but its influence on the interaction between acoustic resonator and reed is not negligible on the threshold frequency, as it is studied now.
\par In a first step, this effect is considered separately, all losses being ignored ($q_r=0$ and $\eta=0$). Equation~(\ref{eq:debitanche}) coupled to Eq.~(\ref{eq:careq}) leads to the following system:
\begin{align}
&-\imag{\mathcal{Y}_e}=k_r\Delta l\frac{\theta}{1-\theta^2},
\label{eq:daimag}\\
&\frac{1}{1-\theta^2}-\frac{1-\gamma}{2\gamma}=0
\Leftrightarrow \gamma=\frac{1-\theta^2}{3-\theta^2}.
\label{eq:dareal}
\end{align}
As seen previously, several frequency solutions $\theta$ exist for different pressure thresholds $\gamma$. Examination of Eq.~(\ref{eq:dareal}) as a function $\gamma=f(\theta)$ (for $\theta<1$) reveals that the solution having the lowest threshold is the one being the closest to the reed frequency.
\par
\begin{figure}
\includegraphics[width=\columnwidth]{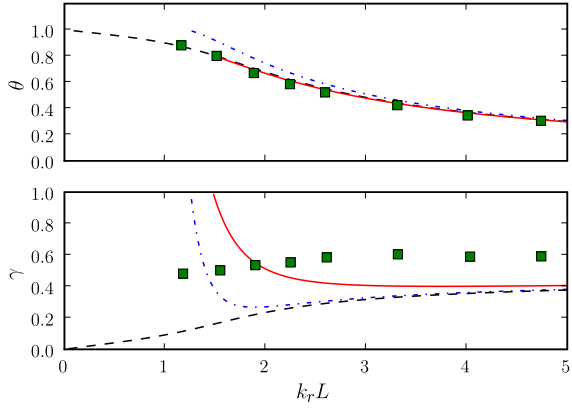}
\caption{Comparing reed motion induced flow effect and reed damping effect ($\Delta l=12\,\milli\meter$ and other conditions as in Fig.~\ref{fig:amortfort1}): reed motion induced flow only (dashed lines),  reed damping only (dash-dotted lines), and both effects (plain lines). WB experimental results are recalled (squares).}
\label{fig:debitanche5}
\end{figure}

\begin{figure}
\includegraphics[width=\columnwidth]{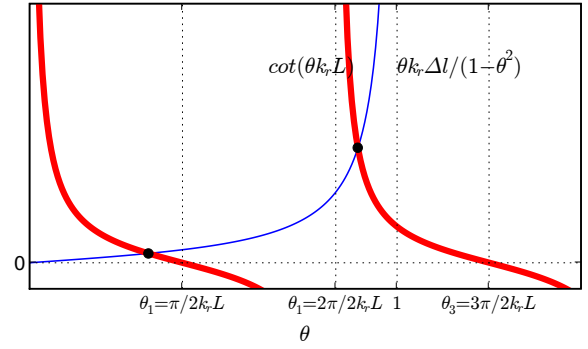}
\caption{Graphical representation of Eq.~(\ref{eq:daimag}) giving oscillation frequency at threshold: left-hand term $-\imag{\mathcal{Y}}$ (thick lines), right-hand term $k_r \Delta \,l\theta/(1-\theta^2)$ (thin line), and solutions (markers).}
\label{fig:dagamma}
\end{figure}

 Approximations can be derived in some situations. When playing close to a bore resonance frequency $\theta_n\ll 1$, the right-hand term in Eq.~(\ref{eq:daimag}) is small, so that, with $\mathcal{Y}_e=-j\,\cot{(\theta k_r L)}$, the reed motion induced flow acts merely as a length correction:
\begin{equation}
\Delta l \frac{1}{1-\theta_n^2}
\mbox{ where }\theta_n=\frac{(2n-1)\pi}{2k_r L}.
\end{equation}
This approximation is valid when the considered bore frequency $\theta_n$ remains smaller than unity. An equivalent approximated length correction can be derived for the effect of reed damping on frequencies from Eq.~(\ref{eq:sys1a}):
\begin{equation}
\Delta l_{q} \simeq \zeta\frac{q_r}{\sqrt{3}k_r}.
\end{equation}
A trial and error procedure has been performed to adjust the oscillation frequency. Very few iterations were needed to exhibit a value for the reed motion induced flow ($\Delta l\simeq\,12\milli\meter$) higher than for the reed damping ($\Delta l_{q}\simeq\,2\milli\meter$) in the conditions of Figs.~\ref{fig:amortfort1} and~\ref{fig:debitanche5}.
When acoustical and mechanical resonances are very close ($\theta=1-\varepsilon$ and $\theta_n=1-\varepsilon_n$), a second-order expression can be deduced:
\begin{equation}
\varepsilon=\frac{\varepsilon_n}{2}
    \left(1+\sqrt{1+2\frac{\Delta l}{L\varepsilon_n^2}}\right),
\end{equation}
the apparition of a square root being typical of mode coupling, making the achievement of analytical expressions difficult. Then, when the bore length decreases enough so that one of its resonances increases above the reed one ($\theta_n>1$), the oscillation frequency approaches the reed one until the intersection point disappears for $k_r L=n\pi $ (see Fig.~\ref{fig:dagamma}). Near the reed resonance, first-order approximations can be derived:
\begin{align}
\theta&\simeq 1-\frac{1}{2}k_r \Delta l\,\tan{(k_r L)},\\
\gamma&\simeq \frac{1}{2}k_r \Delta l\,\tan{(k_r L)}.
\end{align}
These expressions are valid if $\tan{(k_r L)}\gtrsim 0$, i.e., $k_r L\gtrsim n\pi$. According to Eq.~(\ref{eq:dareal}), when oscillation frequency approaches $f_r$, the pressure threshold decreases to zero contrary to what happens when considering the reed damping effect.
\par
\begin{figure}
\includegraphics[width=\columnwidth]{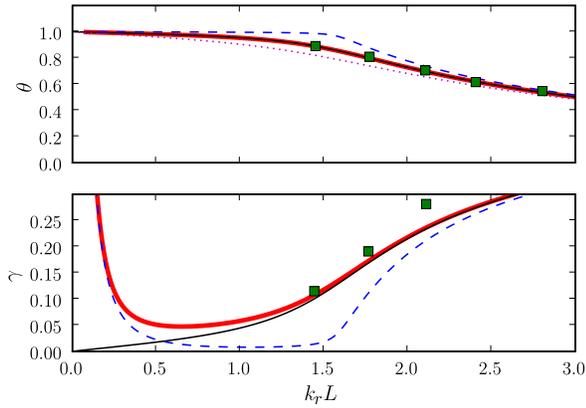}
\caption{Comparing reed motion induced flow effect and reed damping effect (conditions as in Fig.~\ref{fig:amortfaible1}):  reed motion induced flow only ($\Delta l=5\,\milli\meter$: plain thin lines, $\Delta l=12\,\milli\meter$: dotted line in upper graph only), reed damping only (dashed thin lines), and both effects (plain thick lines). WB experimental results are reacalled (squares).}
\label{fig:debitanche2}
\end{figure}

 Figures~\ref{fig:debitanche5} and~\ref{fig:debitanche2} show a numerical comparison of the respective effects of reed motion induced flow and reed damping. The reed motion induced flow with $\Delta l=12\,\milli\meter$ in Fig.~\ref{fig:debitanche5} and $5\,\milli\meter$ in Fig.~\ref{fig:debitanche2} adjusts the frequency deviation for both heavily and slightly damped reeds, even when approaching reed resonance, and is preponderant compared to the damping effect. An example of a mistuned length correction parameter ($\Delta l=12\,\milli\meter$) is exhibited in Fig.~\ref{fig:debitanche2}, revealing a mismatch of oscillation frequency. This latter seems to be mainly controlled by the reed motion induced flow effect when approaching the reed resonance frequency.
Classical approaches (see, e.g., Ref.~~\onlinecite{nederveen:1998}) for the calculation of playing frequencies, ignoring the flow due to pressure drop, and searching for eigenfrequencies of the passive system including the bore and the reed only are thus justified. It can be noticed that Eq.~(\ref{eq:daimag}) was already given by Weber\cite{weber:1828} in the early 19th century (see page 216), assuming a reed area equal to the cylindrical tube section. This theory, used by several authors (see, e.g., Ref.~~\onlinecite{miklos:2006}), was discussed by von~Helmholtz\cite{helmholtz:1954} and Bouasse\cite{bouasse:1930}, especially concerning the lack of explanation on the production of self-sustained oscillations. On the contrary, threshold pressure curves exhibit that both damping and additional flow have influence on the pressure required for the reed to oscillate. So a combination of the phenomena has to be taken into account, none of them being negligible in the considered domain.

\section{Going beyond instability threshold\label{sec:beyondthreshold}}
\subsection{Linear stability analysis with modal decomposition}
\label{subsec:DM}
\begin{figure}[hbt]
\centering
\includegraphics*[width=\columnwidth]{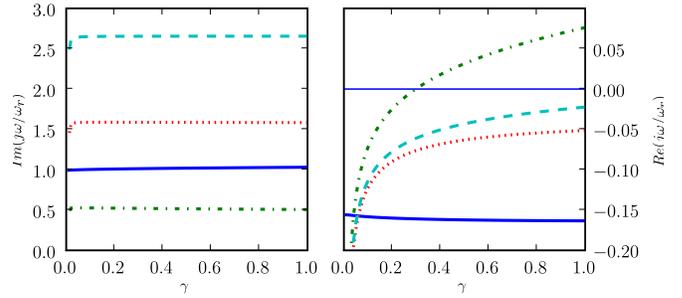}
\caption{Evolution of the complex eigenfrequencies as a function of mouth pressure $\gamma$. $r=7\,\milli\metre$, $\w_r=2\pi\times 1000\,\radian\per\second$, $q_r=0.3$, $L=16\,\centi\metre$ ($\theta_1=0.53$ and $k_r L=2.96$), and $\zeta=0.2$.}
\label{fig:ModesPropres1}
\end{figure}

\begin{figure}[hbt]
\centering
\includegraphics*[width=\columnwidth]{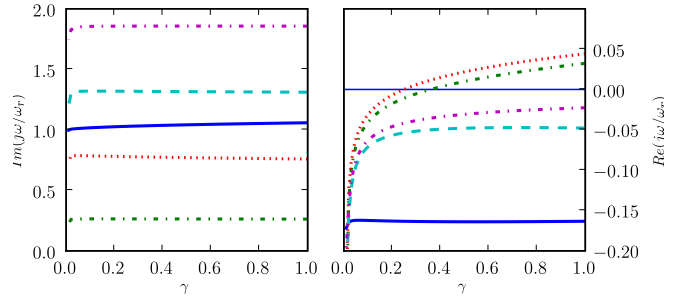}
\caption{Evolution of the complex eigenfrequencies as a function of $\gamma$. $r=7\,\milli\metre$, $\w_r=2\pi\times 1000\,\radian\per\second$, $q_r=0.3$, $L=32\,\centi\metre$ ($\theta_1=0.26$ and $k_r L=5.91$), and $\zeta=0.2$.}
\label{fig:ModesPropres3}
\end{figure}

Analysis of the instability threshold can be performed using complex frequency formalism. For a given configuration of the whole system \emph{bore-reed-musician} ($L$, $S$, $\w_r$, $q_r$, $\gamma$, and $\zeta$ being set), its complex eigenfrequencies $s_n=j\w_n-\alpha_n$ can be determined. The imaginary part of $s_n$ corresponds to the frequency, the real part $\alpha_n$ being the damping of this mode, for the coupled linearized system close to the static equilibrium state. Classically, when the mouth pressure is below the oscillation threshold, all eigendampings $\alpha_n$ are positive, the static regime being stable. An oscillation may appear when at least one mode of the whole system becomes unstable, i.e., when at least one of the $\alpha_n$ becomes negative. Looking for instability threshold can be done by varying a bifurcation parameter (either $L$, $\gamma$, or $\zeta$) and examining the real part of computed eigenfrequencies. Two examples are shown in Figs.~\ref{fig:ModesPropres1} and~\ref{fig:ModesPropres3}. It is noticeable that the frequencies evolve only slightly with the bifurcation parameter $\gamma$ and are close to eigenfrequencies of either the bore or the reed ($\imag{j\w/\w_r}\simeq 1$). For the first example, the static regime becomes unstable for $\gamma\simeq 0.28$ and a frequency near the first resonance of the bore (dot-dashed curves). Other acoustic resonances (their frequencies being odd multiples of the first one) and reed resonance have higher oscillation threshold and remain damped for this configuration. For a longer tube (Fig.~\ref{fig:ModesPropres3}), instability appears for $\gamma\simeq 0.3$ at a frequency located near the third resonance of the bore [dotted line at $\imag{j\w/\w_r}\simeq 0.8\simeq 3\times 0.27$], the first resonance becoming unstable for a larger mouth pressure. The frequency close to the reed resonance (solid curves) still remains damped.
\par Calculations may be simplified and accelerated by using a modal decomposition of the bore impedance $\mathcal{Z}_e(\w)$: this allows for the characteristic equation to be written as a polynomial expression of $j\w$, and optimized algorithms for polynomial root finding can be used. Modal expansion considers the $N$ first acoustic resonances of the bore:
\begin{eqnarray}
\frac{\mathcal{Z}_e(\w)}{Z_c} &=&
    j\,\tan{\left(\frac{\w L}{c}-j\alpha (\w) L\right)}\label{eq:tan}\\
\frac{\mathcal{Z}_e(\w)}{Z_c} &\simeq & \displaystyle
    \frac{2c}{L}\sum_{n=1}^N \frac{j\w}{\w_n^2+jq_n\w\w_n +(j\w)^2},
\end{eqnarray}
where modal coefficients $\w_n$ and $q_n$ can be deduced either from measured input impedance or from analytical expression [Eq.~(\ref{eq:tan})], assuming $\alpha(\w)$ to be a slowly varying function of the frequency.
\par Comparison between direct calculation of oscillation threshold $\gamma_\text{th}$ using WB method and estimation using modal decomposition and complex eigenfrequencies computing has been done. Whereas the difficulties  for the first method arise due to the transcendental characteristic equation, the second one requires calculation of eigenvalues for various values of the mouth pressure $\gamma$, using an iterative search of the instability threshold. The number of modes taken into account has been chosen such that resolution of $\gamma_\text{th}$ is less than $0.01$, which is also the tolerance used for the iterative search. An example is given in Table~\ref{tab:comp} for a heavily damped and strong reed. It shows very good agreement between the two methods for both $\gamma$ and $\theta$. This validation allows the use of the complex frequency approach, which results in an efficient algorithm that can be easily applied to more complex resonators whenever a modal description is available.
\begin{table}[hbt]
\begin{center}
\begin{tabular}[t]{|c|*{2}{|ll|l|}}
\hline
$k_r L$ & $\theta_{CF}$ & $\theta_{WB}$ & $\Delta \theta$
    & $\gamma_{CF}$ & $\gamma_{WB}$ & $\Delta \gamma$\tabularnewline
\hline\hline
8.5 & 0.185 & 0.186 & 0.5\% & 0.43 & 0.43 & 0.1\% \tabularnewline
2 & 0.747 & 0.747 & 0.1\% & 0.30 & 0.30 & 0.6\% \tabularnewline
1 & 1.022 & 1.024 & 0.2\% & 3.86 & 3.82 & 1.0\% \tabularnewline
0.81 & 1.033 & 1.034 & 0.1\% & 8.77 & 8.71 & 0.7\% \tabularnewline
\hline
\end{tabular}
\caption{Comparison of pressure threshold and oscillation frequency calculated using complex frequency formalism (indexed by $CF$) and Wilson \& Beavers method ($WB$) for $r=7\milli\metre$, $\w_r=2\pi\times 750\radian\per\second$, $q_r=0.4$ and $\zeta=0.13$.}
\label{tab:comp}
\end{center}
\end{table}
\par The writing of the characteristic equation for a single acoustic mode exhibits the behavior of the coupled oscillators:
\begin{multline}
\left[\w_n^2 +j\w\left(q_n\w_n+\frac{c}{L}\zeta\frac{1-\gamma}{\sqrt{\gamma}}\right)-\w^2\right]\\
\times\left[\w_r^2+jq_r\w\w_r-\w^2\right]
=j\w\frac{2c}{L}\w_r^2\left(\zeta\sqrt{\gamma}+j\w\frac{\Delta l}{c}\right).
\end{multline}
Coupling realized by the flow in the reed channel modifies the damping of the acoustic mode: in addition to the usual term (corresponding to viscothermal losses and eventually radiation), damping is increased by a quantity related to mouth pressure and stress on the reed. This may be regarded to as a resistive acoustic behavior at the bore entrance.
\par Assuming that a linearized model is still relevant during the growth of oscillations (before the amplitude saturation mechanism appears), this approach can be extended to investigate the transient response of the coupled system. Characterizing the degree of instability of the system by $\sigma=\min_n{\alpha_n}$, the slope of the curve  $\sigma=f(\gamma)$ gives an information about the instability degree of the system when the mouth pressure is slightly higher than the oscillation threshold. A great slope would correspond to a very unstable configuration and a quick growth of oscillation, whereas nearly constant curve would lead to a small amplification coefficient and slowly rising vibrations and then to longer transient attack before stabilization of the magnitude of oscillations. Links between the computed eigenfrequencies of the coupled system presented here and the transient behavior have still to be investigated.

\subsection{Mouth pressure required to obtain a given (small) amplitude}
\label{sec:harmbal}
The previous sections of the paper deal with the stability of the static regime, looking for the condition to make a bifurcation possible. Some developments concerning the existence of oscillating regime above the threshold are derived now. Neither stability of oscillation nor tone deviation issue will be discussed here.
\par Grand \etal\cite{grand:1997} suggested the introduction of the limited Fourier series of pressure in the massless reed case. The technique is the harmonic balance applied to oscillations of small amplitudes. Calculations are done hereafter by taking into account the reed dynamics in the volume flow relationship, which does not appear in the mentioned paper. Fourier series of the volume flow depends on Fourier components  of signal $P_m-p(t)$ and $y(t)$. Assuming steady state oscillations with angular frequency $\w$, the dimensionless signals are written as
\begin{equation}
p(t)=\sum_{n\neq 0} p_n \E{n},\qquad
u(t)= u_0 + \sum_{n\neq 0} \mathcal{Y}_n p_n \E{n},\\
\end{equation}
\begin{equation}
\frac{y(t)}{y_0}=\left(1-\gamma\right)+\sum_{n\neq 0} D_n p_n \E{n},
\end{equation}
where $\mathcal{Y}_n=\mathcal{Y}_e(n\w)$ and $D_n=D(n\w)$ are the values of dimensionless bore admittance and reed transfer function for angular frequency $n\w$. The volume flow relationship is rewritten as
\begin{equation}
u^2(t)=\zeta^2 \left(y(t)/y_0 \right)^2 (\gamma-p(t) ).
\label{eq:Bernoulli}
\end{equation}
Sustained oscillations of very small amplitude are studied, assuming that $p_1$ is a non vanishing coefficient considered as a first-order quantity. Notations $\CE{n}$ and $\F{n}{m}$ are introduced:
\begin{gather}
\CE{n}=\mathcal{Y}_n/(\zeta\sqrt{\gamma})+\frac{1-\gamma}{2\gamma}-D_n,\\
\F{m}{n}=\F{n}{m}=D_n D_m-\frac{1-\gamma}{\gamma}(D_n+D_m)-\frac{\mathcal{Y}_n\mathcal{Y}_m}{\zeta^2\gamma}
\end{gather}
Cancellation of $\CE{n}$ for given $\w$ and $\gamma$ means that the characteristic equation~(\ref{eq:careq}) is solved for $n\w$ and $\gamma$. Expanding Eq.~(\ref{eq:Bernoulli}) leads to
\begin{equation}
\begin{split}
\label{eq:pngeneric}
0 =&\left[\frac{u_0^2}{\zeta^2\gamma}-(1-\gamma)^2 \right]\\
&+2(1-\gamma)\sum_{n\neq 0} \left[
	\frac{u_0\mathcal{Y}_n}{\zeta^2\gamma(1-\gamma)}
	-D_n+\frac{1-\gamma}{2\gamma}\right]p_n \E{n}\\
&-\sum_{n,m\neq 0}\F{n}{m} p_n p_m \E{(n+m)}\\
&+\frac{1}{\gamma}\sum_{n,m,q\neq 0} D_n D_m p_n p_m p_q \E{(n+m+q)}.
\end{split}
\end{equation}
It is here assumed that $p_n$ is of order $|n|$ (with $p_{-n}=p_n^*$) (see, e.g., Ref.~~\onlinecite{grand:1997}). The continuous component of the volume flow is calculated up to order 2:
\begin{equation}
\begin{split}
\frac{u_0^2}{\zeta^2\gamma} = &(1-\gamma )^2 +\sum_{n\neq 0} \F{+n}{-n} | p_n |^2\\
 &-\frac{1}{\gamma} \sum_{n,m,n+m\neq 0} D_n D_m p_n p_m p_{n+m}^*\\
\simeq&  (1-\gamma)^2+2\F{+1}{-1}| p_1 |^2+o(p_1^2)
\end{split}
\end{equation}
so that, considering $u_0$ to be real
\begin{equation}
u_0\simeq \zeta\sqrt{\gamma}(1-\gamma)
\left[1+\frac{| p_1 |^2 \F{+1}{-1}}{(1-\gamma)^2}\right]+o(p_1^2).
\label{eq:u0}
\end{equation}
\par From Eq.~(\ref{eq:pngeneric}), the frequency $(N\w)$ is extracted for $N\geq 1$:
\begin{multline}
0=2(1-\gamma)\left[D_N-\frac{1-\gamma}{2\gamma}-\frac{u_0\mathcal{Y}_n}{\zeta^2\gamma(1-\gamma)}\right]p_N\\
+\sum_{n\neq 0} \F{n}{N-n} p_np_{N-n}\\
-\frac{1}{\gamma}\sum_{n,m\neq 0} D_nD_m p_np_mp_{N-n-m}.
\end{multline}
For $N\geq 2$, Taylor series expansion up to order $N$ is applied: in the first sum, only terms corresponding to $0\leq n\leq N$ contribute at order $N$, while, in the second one, the terms to consider are the ones for which $0<n<N$ and $0<m<N-n$. The component $p_N$ can be deduced from the sequence $( p_n )_{0<n<N}$:
\begin{multline}
p_N=\frac{1}{2(1-\gamma)\CE{N}}\Bigg[
\sum_{0<n<N} \F{n}{N-n} p_np_{N-n}\\
-\frac{1}{\gamma}\sum_{\substack{0<n<N\\0<m<N-n}} D_n D_m p_n p_m p_{N-n-m}\Bigg]+o(p_1^N).
\label{eq:pn}
\end{multline}
As an example, for $N=2$, the second sum being empty:
\begin{equation}
p_2\simeq \frac{\F{1}{1}p_1^2}{2(1-\gamma)\CE{2}}+o(p_1^2).
\end{equation}
As expected and in agreement with the so-called ``Worman rule''\cite{worman:these}, higher components appear to be higher order quantities: order 2 for $p_2$, 3 for $p_3$, 4 for $p_4$, etc.
\par Focus is now given to fundamental frequency. Calculations are done up to order 3:
\begin{equation}
\begin{split}
0 = &2(1-\gamma)
	\left[D_1-\frac{1-\gamma}{2\gamma}-\frac{u_0\mathcal{Y}_1}{\zeta^2\gamma(1-\gamma)}\right]p_1\\
&+\sum_{n \neq 0,1} \F{n}{1-n} p_n p_{1-n}\\
&-\frac{1}{\gamma}\sum_{1-n-m,n,m\neq 0}D_n D_m p_n p_m p_{1-n-m}\\
\simeq & 2(1-\gamma)
	\left[D_1-\frac{1-\gamma}{2\gamma}-\frac{u_0\mathcal{Y}_1}{\zeta^2\gamma(1-\gamma)}\right]p_1\\
&+2\F{2}{-1}p_2p_1^*-\frac{1}{\gamma}D_1p_1|p_1|^2(D_1+2D_1^*)
\end{split}
\end{equation}
Replacing $u_0$ and $p_2$ gives
\begin{equation}
|p_1|^2\simeq \frac{2(1-\gamma )^2 \CE{1}}
{\displaystyle \frac{ \F{1}{1} \F{+2}{-1}}{\CE{2}}
-\frac{2\mathcal{Y}_1}{\zeta\sqrt{\gamma}} \F{+1}{-1}
-\frac{1-\gamma}{\gamma}(D_1^2+2|D_1|^2)}.
\label{eq:p1amp}
\end{equation}
The limitation to the first harmonic method ignoring the influence of $p_{n\geq 2}$ on the amplitude $|p_1|$ leads to
\begin{equation}
|p_1|^2\simeq \frac{2(1-\gamma)^2 \CE{1}}
{\displaystyle -\frac{2\mathcal{Y}_1}{\zeta\sqrt{\gamma}}\F{+1}{-1}-\frac{1-\gamma}{\gamma} (D_1^2+2|D_1|^2)}.
\label{eq:p1ampPH}
\end{equation}

\par In their paper, Grand \etal\cite{grand:1997} stated that the procedure used in a first simplified case can be applied to a model including reed dynamics provided that $\mathcal{Z}_e(\w)$ is replaced by $\mathcal{Z}_e(\w)D(\w)$. Conclusion is not so straightforward as reed dynamics also interferes with the volume flow relationship contributing to a more complex expression for the first component amplitude. According to their work, $D_n$ terms would only occur with $\mathcal{Y}_n$ in the expression of $|p_1|^2$ which is not the case.
\begin{figure}[hbt]
\centering
\includegraphics*[width=\columnwidth]{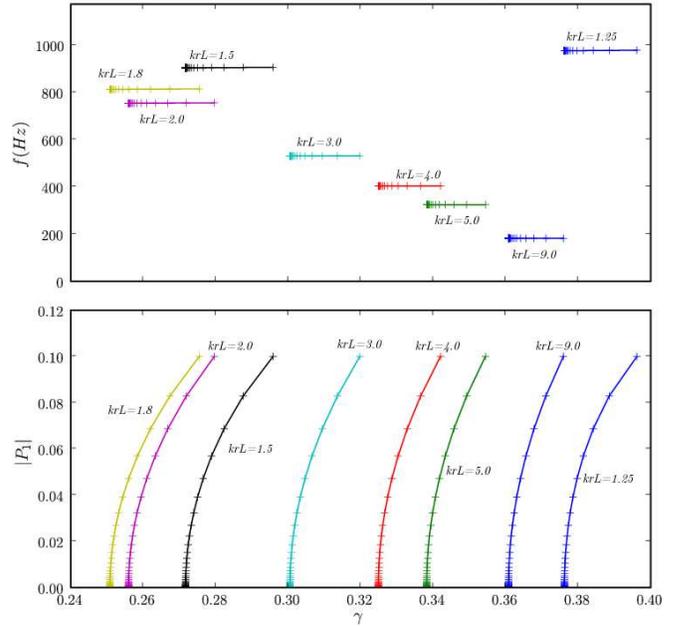}
\caption{Comparison of $|p_1|$ computed with the harmonic balance for small oscillations for various values of $k_r L$: from thin to thick lines, $k_r L=9,\, 5,\, 4,\, 3,\, 2,\, 1.8,\, 1.5,\, 1.25$ ($\Delta l=0$, $q_r=0.4$, and $f_r=1050\,\hertz$).}
\label{fig:BifDirecte}
\end{figure}
\begin{figure}[hbt]
\centering
\includegraphics*[width=\columnwidth]{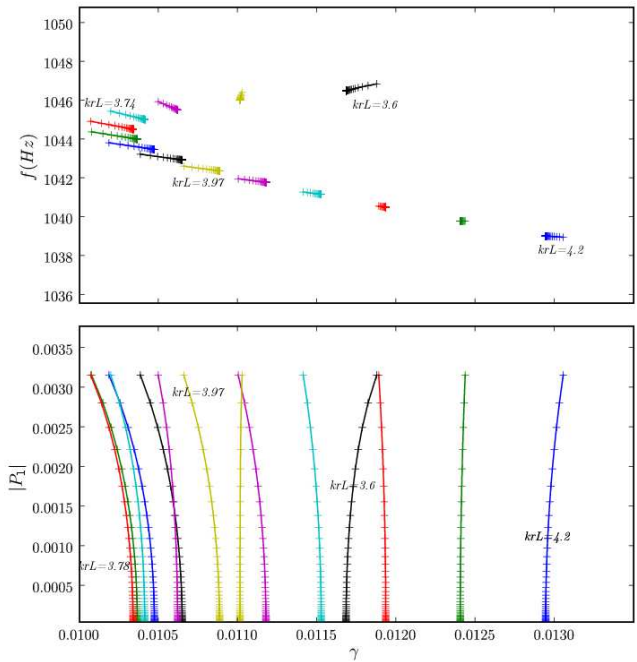}
\caption{Comparison of $|p_1|$ computed with the harmonic balance for small oscillations for 14 values of $k_r L$ regularly spaced between $3.6$ and $4.20$ with a lightly damped reed ($\Delta l=0$, $q_r=0.01$, and $f_r=1050\,\hertz$).}
\label{fig:BifInverse}
\end{figure}
\par In Fig.~\ref{fig:BifDirecte} are shown the bifurcation diagrams for a heavily damped reed. From the flattest tone ($k_r L=9.00$) to the sharpest computed ($k_r L=1.25$), oscillations with very small amplitudes are possible as the diagrams show Hopf bifurcations that are supercritical, i.e., direct, for all the computed cases. On the other hand, for cases where oscillation frequency is closed to the reed one (Fig.~\ref{fig:BifInverse}), a subcritical Hopf bifurcation occurs for a  range of bore length, with a very small pressure threshold linked to the interaction of reed resonance with the second bore resonance: the bifurcation is there inverse.
\par In accordance with Grand \etal\cite{grand:1997}, it appears that the bifurcation is not always direct. There exist configurations for which computed bifurcation diagram shows a subcritical pitchfork, i.e., small oscillations for values of pressure below static regime instability threshold. The boundary between the two cases is not trivial to explore analytically, requiring the mathematical study of Eq.~(\ref{eq:p1amp}). Alternately numerical exploration of parameter space can lead to some partial observations. One of them is that the bifurcation seems to be direct when the oscillation frequency is close to one of the bore resonances.

\par The limit of calculations derived here needs to be emphasized. When $|p_1|$ tends to zero, the oscillation frequency is the one for which the characteristic equation is solved ($\CE{1}=0$). Higher components (the second one here at the first-order approximation) have a non-negligible influence only if $\CE{2}$ becomes small too when approaching oscillation threshold, i.e., if the characteristic equation accepts the solutions $\w$ and $2\w$ for two close values of mouth pressure: $\CE{1}$ and $\CE{2}$ are small simultaneously. Inverse bifurcation may occur in the degenerate case of simultaneously destabilization of static regime for a frequency and its octave. This consideration can be extended to the more general case of $\CE{1}$ and $\CE{N}$ ($N\geq 2$) canceling an equal mouth pressure value.
\par As a conclusion, the nature of the bifurcation depends on the roots of the characteristic equation~(\ref{eq:careq}): this generalizes the results of Grand \etal\cite{grand:1997}.

\section{Conclusion and perspectives}
\label{sec:conlusion}
\par Two components of the volume velocity at the input of the resonator act on the oscillation threshold values: the first one is due to the pressure drop between mouth and mouthpiece, while the second one is due to the reed movement. Roughly speaking, the first one  has a mainly resistive effect, either passive or active, the second one having a mainly reactive effect. This remark can be related to the behavior of the threshold pressures and frequencies. Concerning the frequencies, the second effect is preponderant, at least for the cases studied by Wilson and Beavers, and it can justify the historical method due to Weber\cite{weber:1828} regarding the playing frequencies as eigenfrequencies of a passive resonator.
Concerning the pressure thresholds, the flow due to the pressure drop is essential, and the model based on Bernoulli equation used by Wilson and Beavers\cite{wilson:1974} seems to be satisfactory, but taking into account the flow due to reed movement is necessary, and improves the results of the authors, mainly for lightly damped reeds.
\par Discrepancies between Wilson and Beavers' experimental results and numerical ones still remain. Can  this be due to the nature of the bifurcation at threshold? For a clarinetlike functioning, i.e., for strongly damped reeds, numerical calculation of small oscillations proves that it is supercritical, confirming the works ignoring reed dynamics, while it is not sure for lightly damped reeds. Further experimental investigations are planned to conclude on the nature of the bifurcation on real clarinets. Another unknown concerns the acoustic impedance of the air supply device and of the mouth cavity.
\par Concerning the possible generalization of this work, other kinds of either resonators or reeds should be studied, as, e.g., in Ref.~~\onlinecite{tarnopolsky:2000}. Some authors of the present paper recently obtained simplified theoretical results for an outward striking reed by using a similar method of investigation based on a single degree of freedom reed model\cite{silva:2007c}. Calculations with several DOFs for the reed remain to do, especially for lip reed instruments (see Ref.~~\onlinecite{cullen:2000}), even if recent works dedicated to vocal folds have been done (see Ref.~~\onlinecite{ruty:these}). Finally, the study of transient sounds can be made easier with a good knowledge of the linearized functioning of reed instruments.

\begin{acknowledgments}
The study presented in this paper was led with the support of the French National Research Agency (\textsc{anr}) within the \textsc{consonnes} project. The authors are grateful to A. Hirschberg, Ph. Guillemain, and B. Ricaud for useful discussion.
\end{acknowledgments}

\bibliography{ArticleCompletBib}
\end{document}